\documentclass[10pt,twocolumn]{article}
 
\usepackage[utf8]{inputenc}
\usepackage[T1]{fontenc}
\usepackage{lmodern}
\usepackage[margin=1in]{geometry}
\usepackage{amsmath,amssymb}
\usepackage{graphicx}
\usepackage{booktabs}
\usepackage{hyperref}
\usepackage{cleveref}
\usepackage{microtype}
\usepackage{xcolor}
\usepackage{caption}
\usepackage{subcaption}
\usepackage{float}
\usepackage[numbers,sort&compress]{natbib}
\usepackage{xurl}

\title{%
  TokaMind for Power Grid: Cross-Domain Transfer \\
  from Fusion Plasma\\
  
}

\author{
    JC WU$^{1, \dagger}$, 
    Norton Lee$^{2, *}$, 
    Kai Siang Chen$^{3}$ \\[1ex]
    \small $^{1}$TaiScience Research Group, affiliated with Fu Jen Catholic University, Taiwan \\
    \small $^{2}$Center for Geometry and Physics, Institute for Basic Science (IBS), South Korea \\
    \small $^{3}$Fu Jen Catholic University, Taiwan \\
    \small \texttt{jcwu@taiscience.org}$^{\dagger}$, \texttt{norton.lee@ibs.re.kr}$^{*}$
}

\date{%
    \small \today
}

\begin{document}
\maketitle
 
% ── Abstract ──────────────────────────────────────────────────────────────────
\begin{abstract}
TokaMind \citep{boschi2026tokamind} is a multi-modal
transformer (MMT) foundation model pre-trained on
tokamak plasma diagnostics data from MAST\citep{jackson2024fairmast}, where it
was shown to outperform CNN-based approaches on
fusion benchmarks.
We investigate whether its learned representations generalize to physically distinct but structurally analogous domains.
Through systematic experimentation across four domains-industrial bearing degradation, NASA CMAPSS turbofan degradation, and two independent power grid PMU datasets-we identify four transfer-favoring characteristics that help explain where TokaMind's pretrained representations are most effective:
(1) dense and stable inter-sensor coupling, 
(2) endogenous critical-transition failure modes, 
(3) observed failure occurrence, and 
(4) sufficient labeled events ($N \geq 200$), hereafter referred to as F1--F4. 
Power grid synchrophasor data matches this target-domain profile most directly, while industrial degradation datasets demonstrate that TokaMind can still yield useful performance under partial alignment, especially when task design and feature construction expose physically meaningful degradation structure.
 
On the GESL/PNNL 500-event benchmark with provider-aware evaluation, TokaMind achieves test $\text{F1} = 0.837 \pm 0.040$ (3~seeds) for severe event classification.
Our central finding, however, is not the aggregate score: classification difficulty is structurally determined by provider-level grid topology, not model capacity.
In the single-window early-warning regime (\texttt{seq\_len}=1), TokaMind outperforms a CNN baseline (F1~0.889 vs.~0.878)--a reversal that disappears as more event windows are provided.
Furthermore, Critical Slowing Down (CSD) indicators, used as a confidence gate rather than a classification label, improve F1 from 0.696 to 0.750 at 63\%
coverage-outperforming the CNN baseline (0.636) at any coverage level.
These results establish the first cross-domain validation of TokaMind outside
nuclear fusion and propose a transferability framework and revised evaluation
protocol for multi-source PMU datasets.
\end{abstract}


% ── 1. Introduction ───────────────────────────────────────────────────────────
\section{Introduction}
\label{sec:intro}

Foundation models pre-trained on large corpora have
demonstrated remarkable transfer capabilities across
natural language and vision domains \citep{bommasani2021opportunities}.
Recent work has begun extending this paradigm toward
scientific machine learning and physics-informed
models, where representations are shaped not only by
data but also by underlying physical structure
\citep{karniadakis2021physics,cuomo2022scientific,subramanian2023towards}. TokaMind \citep{boschi2026tokamind} is a compact ($<$10M parameter) multi-modal transformer pre-trained on MAST tokamak diagnostics, using DCT3D tokenization \citep{boussakta2004dct3d} to compress heterogeneous sensor streams into a unified representation.
Its architecture explicitly models inter-sensor coupling at multiple temporal scales, reflecting the underlying magnetohydrodynamic (MHD) constraints of plasma physics.

TokaMind may be viewed as an instance of a multimodal scientific foundation model, where heterogeneous sensor streams are fused into a shared representation space. This aligns with broader efforts in multimodal and generalist learning systems that aim to unify representations across modalities \citep{reed2022generalist}. We ask: does TokaMind's learned representation of physically-coupled multi-sensor dynamics transfer to other domains governed by analogous physical constraints?
Power grid synchrophasor (PMU) measurements provide high-resolution, time-synchronized observations of system dynamics and have become a standard tool for wide-area monitoring and stability analysis \citep{phadke2017synchronized,biswas2023open,kundur2004definition,meng2019fast}.
They are governed by Kirchhoff’s circuit laws—a fixed physical constraint structurally analogous to MHD.
Grid disturbance events correspond to dynamical instability phenomena such as voltage collapse and frequency excursions \citep{kundur2004definition,ornl2023ges}, representing genuine phase transitions in a dynamical system rather than gradual degradation.

\begin{figure}[H]
  \centering
  \includegraphics[width=\columnwidth]{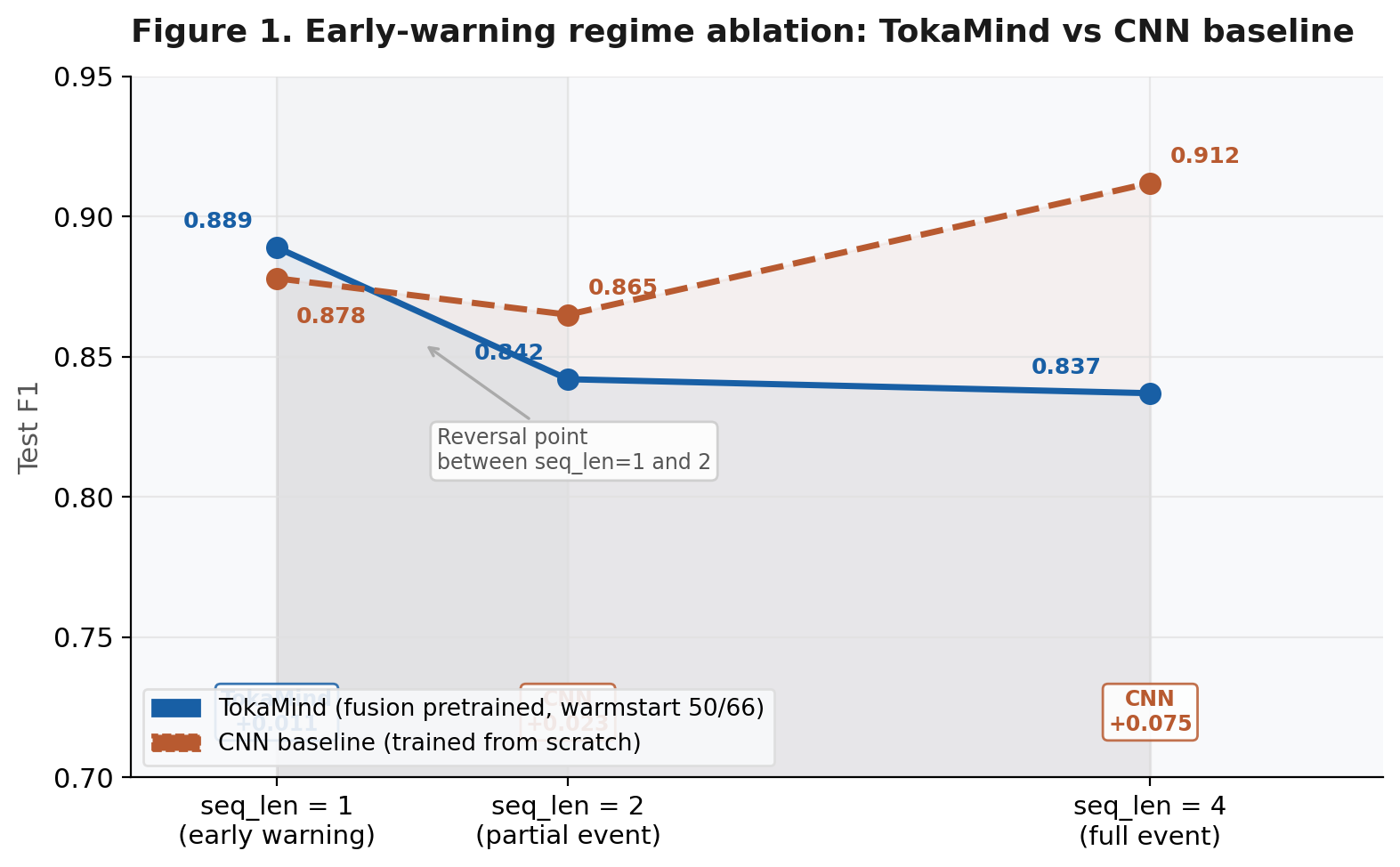}
    \caption{Test F1 vs.\ number of input windows (\texttt{seq\_len}).
           TokaMind leads in the single-window early-warning regime;
           CNN recovers as more windows are provided.
           Reversal point between \texttt{seq\_len}=1 and \texttt{seq\_len}=2.}
  \label{fig:seqlen}
\end{figure}

\begin{figure}[H]
  \centering
  \includegraphics[width=\columnwidth]{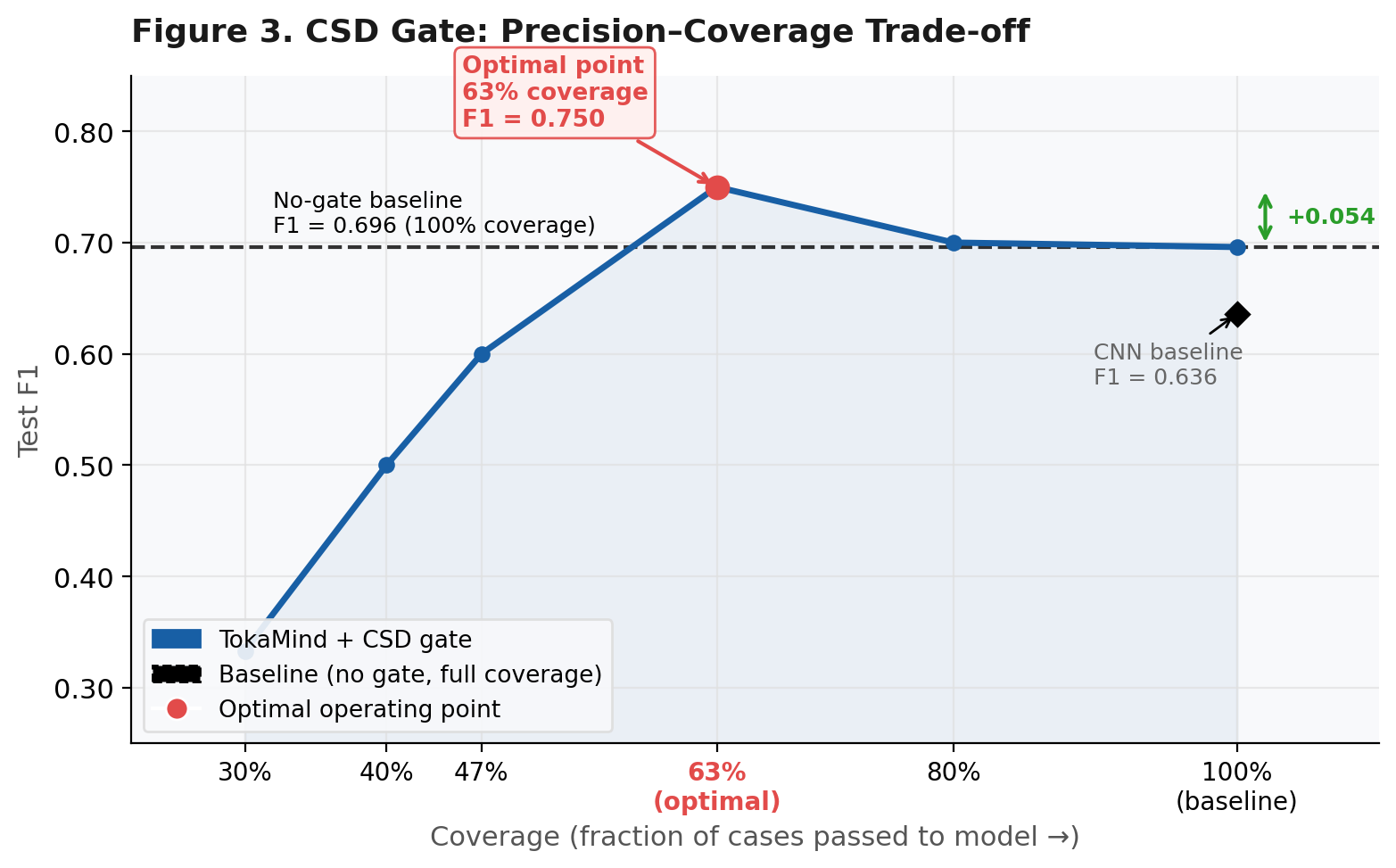}
    \caption{CSD selective prediction framework.
           TokaMind inference and CSD computation run in parallel.
           }
  \label{fig:csdflow}
\end{figure}

\begin{figure}[H]
  \centering
  \includegraphics[width=\columnwidth]{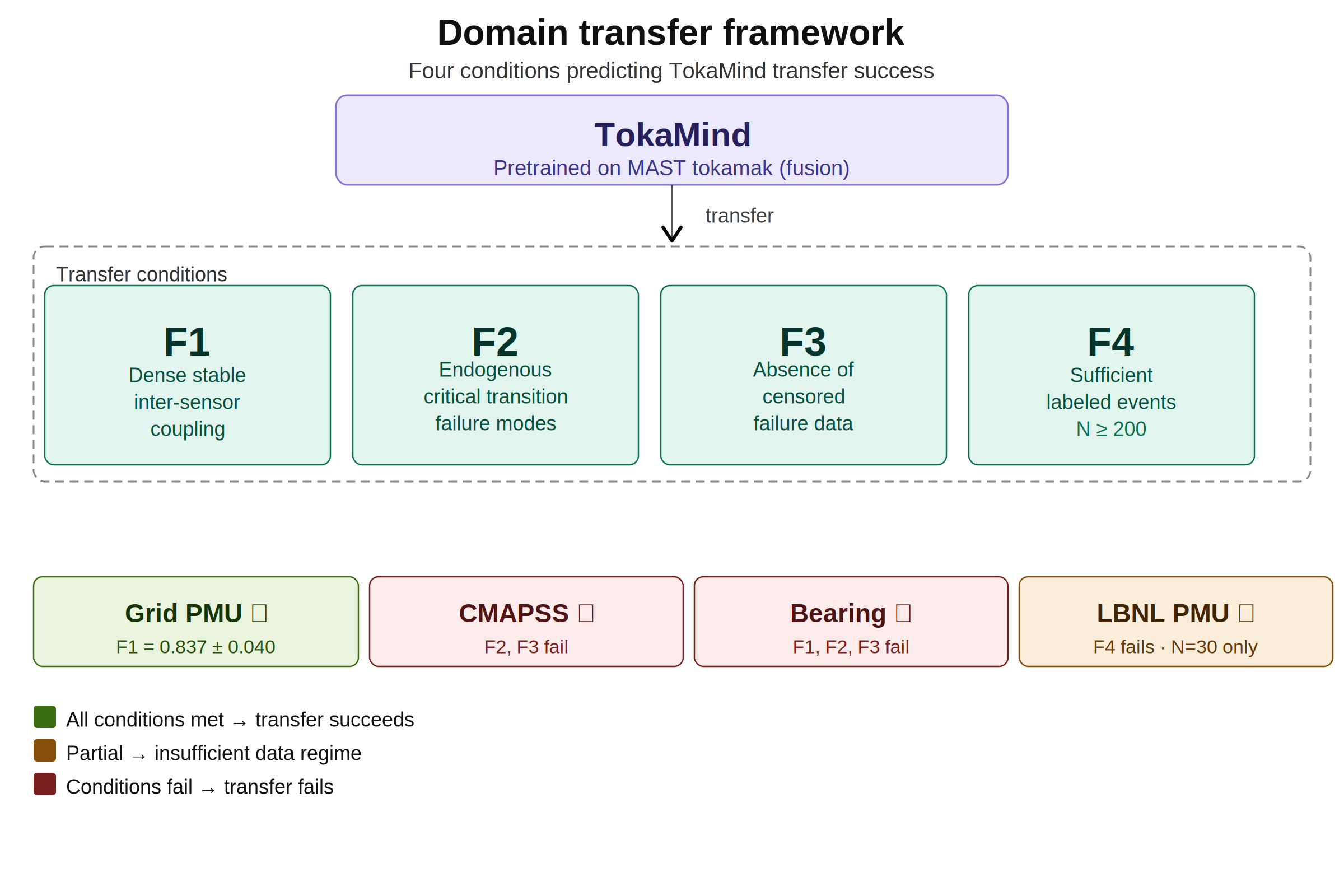}
  \caption{Transfer-favoring characteristics (F1–F4) associated with successful cross-domain transfer of TokaMind, evaluated across Grid PMU, CMAPSS, Bearing, and LBNL PMU datasets.}
  \label{fig:framework}
\end{figure}

Our contributions are:
\begin{enumerate}
  \item \textbf{Systematic transfer behavior analysis.}
        We evaluate TokaMind across domains with
        different degrees of physical alignment, including bearing degradation, CMAPSS,
        small-sample PMU, and GESL/PNNL PMU data. Rather than treating unsuccessful settings as simple failures, we use them to derive a practical target-domain
profile for future TokaMind applications.
  \item \textbf{Successful cross-domain transfer.}
        TokaMind achieves $\text{F1} = 0.837 \pm 0.040$ on GESL/PNNL
        under rigorous provider-aware evaluation.
  \item \textbf{Early-warning regime reversal.}
        At \texttt{seq\_len}=1, TokaMind outperforms CNN (0.889 vs.\ 0.878);
        the advantage reverses at \texttt{seq\_len}=4.
        See \cref{fig:seqlen}.
  \item \textbf{Provider-level observability finding.}
        Classification difficulty is structurally determined by grid topology,
        not model capacity.
        See \cref{fig:provider}.
  \item \textbf{CSD as selective prediction gate.}
        CSD indicators improve F1 from 0.696 to 0.750 at 63\% coverage.
        See \cref{fig:csdflow}.
  \item \textbf{A transferability framework.}
        Four physical conditions predicting when TokaMind-style models transfer successfully.
        See \cref{fig:framework}.
\end{enumerate}
 
% ── 2. Background ─────────────────────────────────────────────────────────────
\section{Background}
\label{sec:background}
 
\subsection{TokaMind Architecture}
TokaMind employs a Multi-Modal Transformer (MMT) with DCT3D tokenization \citep{boschi2026tokamind}.
Each modality's time series is transformed via 3D Discrete Cosine Transform into fixed-length tokens (\texttt{token\_dim}=512), enabling processing of signals at different sampling rates.
The pre-training objective learns to predict masked tokens across modalities, implicitly modeling inter-sensor correlations shaped by MHD physics. Evaluation is standardized via TokaMark \citep{rousseau2026tokamark}. TokaMind's four pre-training objectives—equilibrium reconstruction, fast magnetics, profile dynamics, and MHD prediction—collectively foster a deep representation of the system’s state space near critical boundaries. This high-dimensional understanding of continuous physical evolution inherently equips the model to capture the early onset of system instability.
 
\subsection{Failure Observability and Dataset Alignment}
Industrial field data is often subject to censoring, as equipment is replaced before catastrophic failure, resulting in datasets dominated by early and mid-stage degradation \citep{lei2020machinery}. In contrast, laboratory benchmarks such as CMAPSS and bearing test datasets provide run-to-failure
trajectories, but their supervised tasks are typically centered on gradual prognostic degradation rather than explicit failure occurrence.

This difference highlights an important alignment issue: TokaMind is pre-trained on tokamak diagnostics data, where multi-channel signals arise from strongly coupled physical processes and often reflect regime-dependent system dynamics. However, many industrial benchmarks emphasize pre-failure prediction without directly modeling failure occurrence as an observable event. 
As a result, domains that expose the onset of physical instability as an observable, continuous phenomenon are more naturally aligned with TokaMind's pretraining bias toward MHD phase transitions, whereas purely prognostic settings may require additional task reformulation or feature design to fully exploit its representations. This positions TokaMind within the emerging class of scientific foundation models for general time-series analysis, aiming to achieve cross-domain generalization in a manner analogous to contemporary large-scale architectures \citep{jin2024timellm,wu2023timesnet}.

\subsection{Critical Slowing Down}

Critical Slowing Down (CSD) is a dynamical phenomenon that arises as a system approaches a critical transition or bifurcation point\citep{lade2012early,boettiger2012early}. As the dominant eigenvalue of the underlying dynamics approaches zero, the system's recovery rate from perturbations decreases, leading to characteristic statistical signatures such as increased lag-1 autocorrelation, rising variance, and enhanced temporal persistence \citep{scheffer2009early,scheffer2012anticipating,kuehn2011mathematical,ditlevsen2010tipping,dakos2012methods}. More generally, critical-transition phenomena have long been associated with anomalous responses in classical physical systems, including variations in sound propagation near phase equilibrium boundaries \citep{novikov1960sound}.CSD has been extensively studied as an early-warning signal across a range of complex systems, including ecological regime shifts \citep{dakos2008slowing,scheffer2009early,dakos2012methods}, climate tipping elements \citep{lenton2012early}, and neurological transitions such as epileptic seizures \citep{meisel2015intrinsic}. In these settings, CSD indicators are typically used to detect proximity to critical transitions, rather than to directly classify system states.

In this work, we adopt a different perspective. Instead of
treating CSD-derived indicators as classification labels,
we use them as a physics-informed confidence signal for
selective prediction, closely related to selective classification
with reject option \citep{chow1969optimum,el2010foundations,geifman2017selective}.
Specifically, we use CSD metrics to identify regions of the input space that are more consistent with endogenous approach-to-instability dynamics, and restrict predictions to these regions. This reframes CSD from an early-warning detector into a gating mechanism that improves robustness and interpretability under cross-domain transfer. As a result, domains that expose failure as an observable, event-level phenomenon are more naturally aligned with TokaMind’s pretraining bias. In power systems, PMU-based disturbance detection and classification has been widely studied using both model-based and data-driven approaches \citep{li2021robust,milano2018lowinertia}.

% ── 3. Failed Transfer Attempts ───────────────────────────────────────────────
\section{Boundary Cases and Partial Alignment}
\label{sec:failures}
 
\subsection{Industrial Bearing Degradation}
We evaluate TokaMind on the FEMTO-ST bearing dataset, which contains real-world accelerated degradation data from factory floor bearings.
Bearing fault signatures are impulsive: rolling element defects produce periodic impulse trains whose time-frequency structure is fundamentally different from the continuous coupled oscillations of plasma or power grid signals.
DCT3D, originally designed for continuous multi-modal fields, may be less naturally aligned with sparse impulsive events.
Inter-sensor coupling is configuration-dependent and not governed by a fixed physical law, unlike MHD or physical coupling structures in multi-sensor dynamical systems.

Furthermore, factory maintenance practice introduces censored data: bearings are replaced before catastrophic failure. This preventive cutoff truncates the signals exactly when the onset of critical instability would begin, preventing the model from observing the continuous precursor dynamics its pre-trained representations are sensitive to.
TokaMind demonstrates a clear transfer failure on the FEMTO-ST dataset, confirming that domains lacking the identified favorable factors (F1–F3) are not currently suitable for direct transfer from the fusion domain.

\subsection{NASA CMAPSS Turbofan Degradation}
NASA CMAPSS consists of multivariate turbofan sensor trajectories simulated under different operating conditions and fault model
\citep{saxena2008cmapss}.
 While CMAPSS provides ground-truth failure labels, these are typically used as terminal points for Remaining Useful Life (RUL) regression. This task formulation focuses on the statistical distance to a predefined end-state rather than the detection of a discrete physical transition onset. In our framework, CMAPSS does not exhibit the favoring characteristic F3 because its `failure' is a cumulative degradation threshold, not the kind of endogenous, abrupt phase transition that TokaMind's representations-learned from magnetohydrodynamic (MHD) instabilities-are naturally sensitive to. Furthermore, because its sensor relationships are conditioned by shifting operating regimes rather than governed by a dense, stable physical coupling law, the dataset also lacks favoring characteristics F1 and F2, though it aligns with the favoring characteristic F4 in terms of data scale.

\subsection{LBNL PMU Event Library (Insufficient Data)}
The LBNL PMU Event Library provides high-resolution synchrophasor measurements of real-world grid anomalies. As a continuous, physics-governed system experiencing discrete faults, it successfully exhibits favoring characteristics F1, F2, and F3. However, with only $N=30$ recorded events, it lacks sufficient data scale (F4) for a robust fine-tuning evaluation. Under 5-fold cross-validation ($\sim$6 events per fold), reliable threshold calibration and $F_1$ estimates become statistically infeasible. Although a PR-AUC of 0.80 suggests the model learned useful precursor structures, the limited sample size precludes definitive threshold-based evaluation.

% ── 4. Successful Transfer ────────────────────────────────────────────────────
\section{Successful Transfer: Power Grid PMU Classification}
\label{sec:results}
 
\subsection{Physical Basis for Transfer}
Power grid synchrophasor measurements satisfy all four transfer-favoring characteristics. PMUs provide high-resolution, time-synchronized measurements of system dynamics and are widely used for monitoring and control of large-scale power systems
\citep{phadke2017synchronized}.
Voltage, current, and frequency signals are coupled by structured coupling dynamics in physical multi-sensor systems.
Grid disturbance events correspond to dynamical instability phenomena extensively studied in power
systems literature \citep{kundur2004definition}.

Grid disturbance events represent genuine endogenous phase transitions. Fault events are recorded by PMUs and labeled by grid operators; no preventive censoring occurs.
 
\subsection{Dataset: GESL/PNNL 500-Event Library}
We use the ORNL Grid Science Event Library (GESL), a 500-event subset of the
PNNL open-source PMU library \citep{biswas2023open}, containing
transmission-level synchrophasor measurements from 13 providers across the United States.
 
\textbf{Preprocessing.}
Three-phase voltage sequences are extracted and windowed (window expansion within event), processed via STFT $\rightarrow$ C$\times$F$\times$T
time-frequency cube $\rightarrow$ DCT3D compression to
\texttt{token\_dim}=512, with \texttt{seq\_len}=4.
 
\textbf{Labeling.}
Severity scores are computed from voltage nadir depth, duration, and rate-of-change. Binary labels assigned at the 75th percentile
(\texttt{pos\_ratio}=0.25).
 
\textbf{Split strategy.}
Provider-aware stratified split ensuring all providers represented in
train/val/test. Final split: train/val/test = 346/71/83.
Class weights [0.503, 1.497] applied for imbalance.
 
\subsection{Two-Stage Adaptation Protocol}
\begin{figure}[H]
  \centering
  \includegraphics[width=\columnwidth]{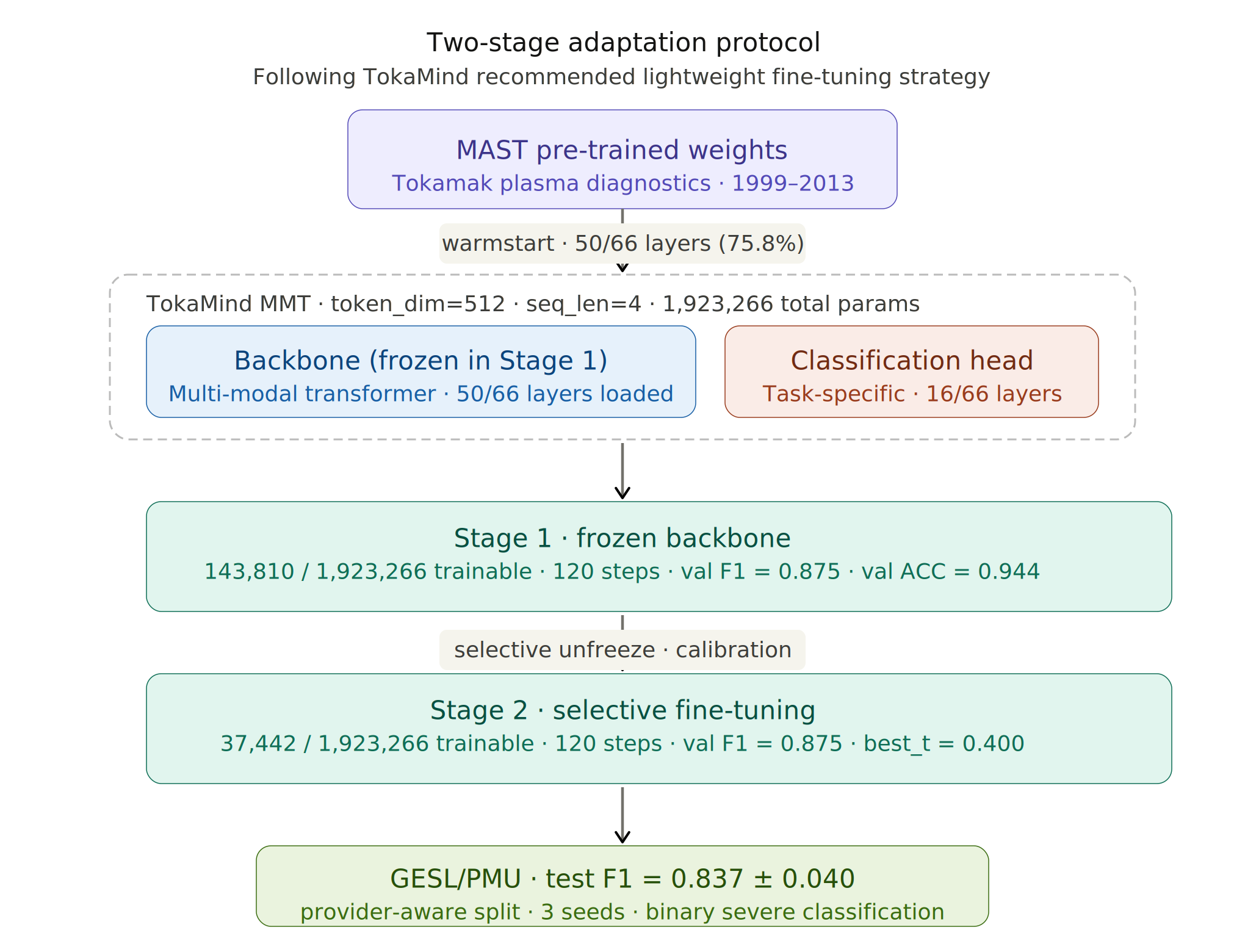}
  
  \caption{Two-stage adaptation protocol following TokaMind.
           MAST pre-trained weights loaded as warmstart (50/66 layers, 75.8\%).
           Stage~1 freezes backbone (143,810 trainable params);
           Stage~2 applies selective fine-tuning (37,442 trainable params).}
  \label{fig:protocol}
\end{figure}

Following the lightweight fine-tuning strategy recommended by TokaMind\cite{boschi2026tokamind}, we load 50/66 pre-trained layers as warmstart
(75.8\%), then apply two-stage training (\cref{fig:protocol}):
 
\begin{itemize}
  \item \textbf{Stage~1} (frozen backbone): 143,810/1,923,266 trainable
        parameters, 120 steps.
        Result: val~F1 = 0.875, val~ACC = 0.944.
  \item \textbf{Stage~2} (selective fine-tuning): 37,442 trainable parameters,
        120 steps.
        Result: val~F1 = 0.875, best threshold = 0.400.
\end{itemize}

Fine-tuning pre-trained components are selectively loaded to preserve transferable representations while minimizing 
catastrophic forgetting. 
Stage~1 establishes the classification boundary using the frozen 
fusion-pretrained backbone; 
Stage~2 refines calibration with minimal parameter updates, 
contributing primarily to probability output 
stability rather than boundary shift. The full 
protocol is illustrated in \cref{fig:protocol}.

\subsection{Main Results}
 
\begin{table}[H]
\centering
\caption{Main results on GESL/PNNL 500-event benchmark.
         Provider-aware split, binary severe/non-severe classification.}
\label{tab:main}
\small
\begin{tabular}{p{3.2cm}cc}
\toprule

Model & test F1 & test ACC \\
\midrule
CNN baseline (seq\_len=4) & $0.912 \pm 0.013$ & $0.960 \pm 0.006$ \\
TokaMind warmstart (seq\_len=4) & $0.837 \pm 0.040$ & $0.924 \pm 0.023$ \\
\midrule
CNN baseline (seq\_len=1) & 0.878 & 0.940 \\
\textbf{TokaMind warmstart (seq\_len=1)} & \textbf{0.889} & \textbf{0.952} \\
\midrule
TokaMind + CSD gate ($\gamma$=0.40) & 0.750 & --- \\
TokaMind + CSD gate ($\gamma$=0.20) & 0.700 & --- \\
CNN baseline (Group~A, seq\_len=4) & 0.636 & 0.775 \\
\bottomrule
\end{tabular}
\end{table}
 
\subsection{Seqlen Ablation: Early-Warning Regime}

At \texttt{seq\_len}=1, TokaMind leads by 0.011 F1 points
(0.889 vs.\ 0.878), consistent with its pre-trained
sensitivity to single-window transition signatures.
The margin reverses at \texttt{seq\_len}=2 (CNN +0.023)
and widens at \texttt{seq\_len}=4 (CNN +0.075), where
CNN's local amplitude aggregation benefits from
accumulated event context.

This result suggests that TokaMind's fusion-pretrained physical coupling representations carry unique value in the information-minimal early-warning setting--precisely where CNN's local amplitude aggregation fails.
 
\subsection{Provider-Level Analysis}
\label{sec:provider}
 
\begin{figure}[tb]
  \centering
  \includegraphics[width=\columnwidth]{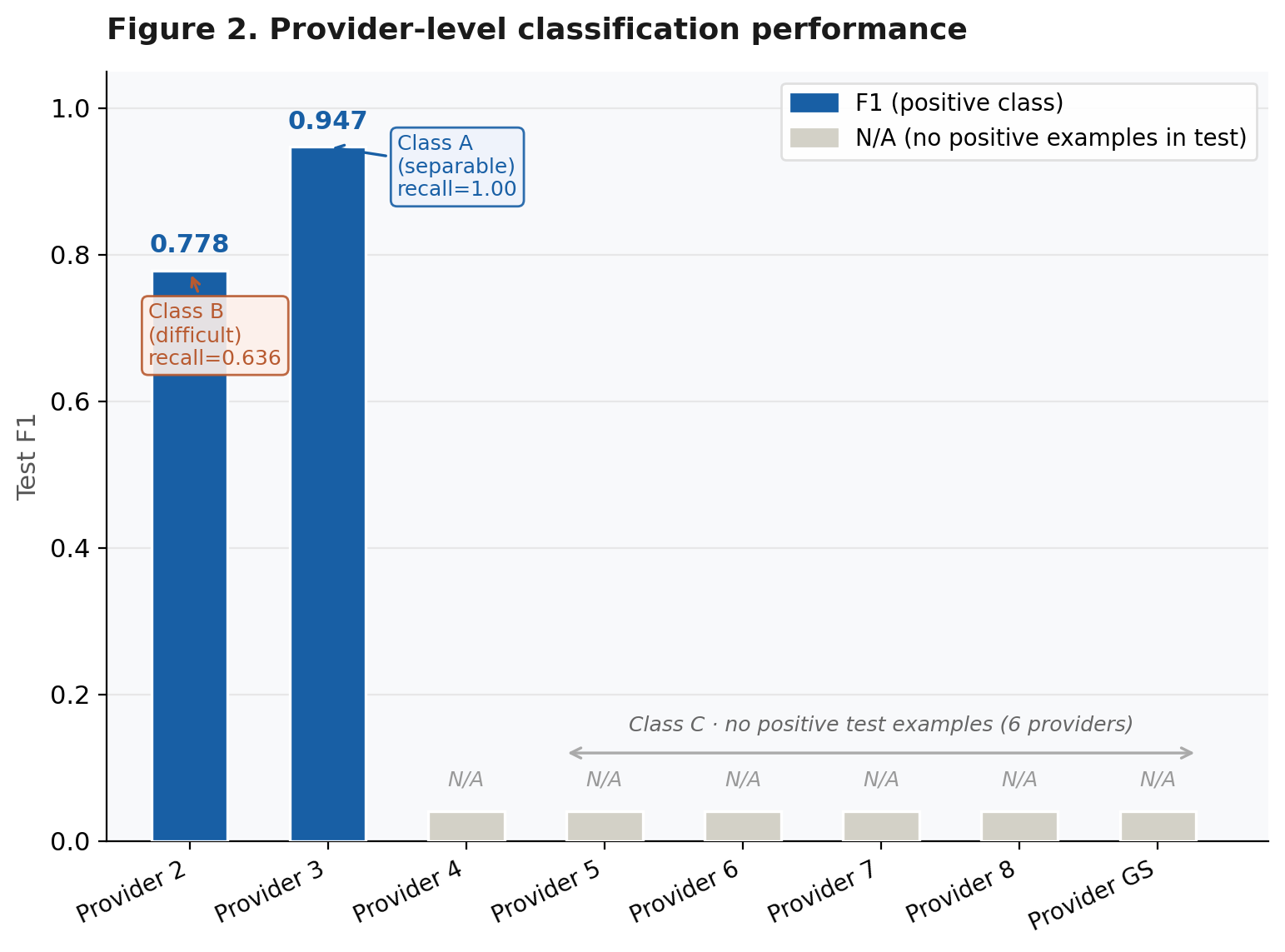}
  
  \caption{Per-provider test F1.
           Classification difficulty is structurally determined by
           provider-level grid topology.
           Overall \texttt{acc}=0.94 is inflated by all-negative providers and should not be interpreted as uniform generalization.}
  \label{fig:provider}
\end{figure}

\Cref{fig:provider} shows per-provider test F1.
Three behavioral classes emerge:
 
\begin{itemize}
  \item \textbf{Class~A (separable):} Provider~3,
        F1~=~0.947, recall~=~1.00.
        Strong unambiguous severe event signatures.
  \item \textbf{Class~B (difficult):} Provider~2,
        F1~=~0.778, recall~=~0.636.
        Conservative prediction; more complex grid topology.
  \item \textbf{Class~C (unobservable):} Remaining providers.
        No positive test examples under global severity threshold.
        \texttt{acc}~=~1.00 with F1~=~N/A.
\end{itemize}
 
Provider~4 (113 events, 100\% trip/generator events) exhibits severity score standard deviation of 0.0--a consequence of metadata template homogeneity rather than physical uniformity--rendering its severity labels unreliable.
This demonstrates the need for per-provider label quality auditing in multi-source PMU benchmarks.
 
\textbf{Evaluation recommendation.}
We propose \textit{positive-provider F1}, macro F1, and recall as primary metrics for multi-source PMU classification, replacing overall accuracy.
 
% ── 5. CSD Gate ───────────────────────────────────────────────────────
\section{CSD as Selective Prediction Gate}
\label{sec:csd}

\subsection{CSD Development: From Label to Gate}

We explored three formulations of Critical Slowing Down indicators before arriving at the final design.

\textbf{Version 1 (Metadata severity score).}
The first formulation derived a severity score directly from event metadata in \texttt{parameter.csv}, combining voltage nadir depth, duration, and rate-of-change into a scalar label. This approach is not a true CSD indicator-it quantifies event outcome rather than dynamical proximity to a critical transition. More critically, under provider-aware evaluation, the score distribution exhibited systematic provider-level stratification: Provider~4 produced a constant severity score (std~=~0.0) due to metadata template homogeneity, rendering its labels unreliable. This constitutes a form of provider leakage that inflates apparent classification performance \citep{scheffer2012anticipating}.

\textbf{Version 2 (Onset-aligned lag-1 autocorrelation).}
The second formulation computed lag-1 autocorrelation
(AC) slopes over pre-event background windows,
following the canonical CSD early-warning framework
\citep{scheffer2009early}.
Three implementation limitations degraded performance
to F1\,$\approx$\,0.34.
First, event onset timestamps were not available in
the GESL metadata, preventing precise alignment of
background windows to the pre-transition period.
Second, the window ratio parameter
(\texttt{window\_ratio}\,=\,0.35)produced excessively long windows for events with large total signal length, smoothing out short-term AC trends. Third, provider-level variation in background noise characteristics dominated the AC signal, obscuring event-level criticality.

\textbf{Version 3 (Provider-normalized CSD).}
The third formulation applied global z-score normalization within each provider before computing CSD scores, attempting to remove provider-level baseline shifts. Performance improved marginally to F1~=~0.53 but remained well below the baseline classifier, confirming that the fundamental limitation was not normalization but the absence of reliable onset alignment.

\textbf{Final design (CSD as confidence gate).}
Rather than using CSD as a classification label, we
reframed it as a \textit{confidence gate}. Although
precise onset alignment is unavailable, CSD indicators
still capture useful dynamical regularity in the signal.
Events with higher CSD scores tend to exhibit more
stable TokaMind probability outputs, consistent
with prior work relating calibrated predictive
confidence to uncertainty estimation in deep neural
networks \citep{lakshminarayanan2017simple}.

We therefore use the CSD score not to classify events directly, but to determine which events are classified automatically and which are routed to human review. Specifically, events with
CSD score above threshold $\gamma$ are classified
automatically; others are deferred to human review.

\subsection{Selective Prediction Trade-off and Operating Regime}

The coverage--F1 trade-off across $\gamma$ values demonstrates
that selective prediction strictly dominates full automation,
consistent with classical coverage-risk trade-offs in
selective classification literature
\citep{el2010foundations,geifman2017selective}.

At $\gamma=0.40$ (coverage=63\%), F1 improves from 0.696 to
0.750, outperforming the CNN baseline (0.636) at any coverage
level. At $\gamma=0.20$ (coverage=80\%), F1=0.700 remains
above the CNN baseline with higher automation.

Below 47\% coverage, the retained cases are too few to
maintain meaningful throughput. This defines a practical
operating band of 47\%--80\% coverage, within which the
CSD gate consistently outperforms both the no-gate baseline
and the CNN baseline.

This selective prediction design is operationally viable
for grid protection systems where human-in-the-loop review
is feasible for a minority of events. 

The 37\% routed to human review, aligning with learning-to-defer frameworks where uncertain predictions are delegated to external decision-makers \citep{madras2018predict,pan2010survey}.
Routing these cases to human review therefore improves
both precision and operational safety simultaneously.
 
% ── 6.Transfer-favoring characteristics────────────────────────────────────
\section{Transfer-favoring characteristics}
\label{sec:framework}

Transferability depends on structural alignment between source and target domains, not superficial physical similarity.
The four characteristics (F1--F4) collectively describe
the degree to which TokaMind's pretrained representations
remain informative under domain shift.

The GESL/PNNL grid PMU corpus most closely satisfies the proposed transfer-favoring characteristics, yielding test
F1 $= 0.837 \pm 0.040$; CMAPSS and Bearing fail primarily on F1, F2 and F3. LBNL PMU remains a boundary case, physics-compatible but below the F4 sample
threshold ($N = 30$), and is excluded from the main evaluation.
F1--F4 thus serve as a lightweight pre-screening protocol before any fine-tuning
compute is committed.

In the meantime we adopt TokaMind's recommended lightweight fine-tuning strategy, adapting the pre-trained model to the power grid
PMU domain via a two-stage protocol.

\begin{description}
  \item[Stage 1 --- Frozen backbone.]
        The backbone (50/66 layers) is initialized from MAST
        pre-trained weights and held frozen. Only the
        task-specific classification head (16/66 layers) is
        trained for 120 steps, allowing the head to align
        with PMU feature distributions without disturbing
        the pre-trained representations.
        Validation F1 = 0.875, val ACC = 0.944.

  \item[Stage 2 --- Selective fine-tuning.]
        The backbone is selectively unfrozen and the full
        model is fine-tuned for 120 additional steps with
        a reduced learning rate. This allows domain-specific
        adaptation while preserving the physical coupling
        representations acquired during MAST pre-training.
        Validation F1 = 0.875, best\_t = 0.400.
\end{description}

The final model is evaluated on the GESL/PMU held-out
test set under provider-aware split across 3 seeds,
yielding test F1 = $0.837 \pm 0.040$.

% ── 7. Discussion ─────────────────────────────────────────────────────────────
\section{Discussion}
\label{sec:discussion}
 
\textbf{Different inductive biases across event regimes.}
On mixed datasets with full event sequences, CNN achieves
higher overall performance than TokaMind. We attribute this
to CNN's effectiveness at exploiting localized temporal
patterns and operator-triggered event signatures that are
prominent in long event windows. In contrast, TokaMind
appears comparatively more effective in physically purified
and information-limited regimes, where cross-sensor
coupling structure becomes more important than extended
event-specific statistical cues. After restricting the
evaluation to endogenous phase-transition events only
(Group A), the performance gap between CNN and TokaMind
substantially narrows, suggesting that the two models
capture complementary aspects of the underlying dynamics.
 
\textbf{Provider heterogeneity as physical structure.}
The provider-level F1 distribution reflects genuine physical heterogeneity in grid topology, not model deficiency.
Provider~4's severity score homogeneity (std~=~0.0) reveals a metadata quality problem that corrupts label reliability--a finding applicable to any multi-source physical benchmark.
 
\textbf{CSD as confidence, not label.}
CSD indicators capture dynamical proximity to critical transitions.
Using CSD as a classification label failed under provider-aware evaluation because the signal reflects provider-level background noise characteristics rather than event-level criticality.
As a confidence gate, however, CSD successfully identifies the minority of events where TokaMind's probability output is well-calibrated, improving precision at the cost of coverage.
 
\textbf{Implications for TokaMind.}
Our findings suggest that TokaMind's pre-trained representations encode physically meaningful structure transferable across domains sharing analogous physical coupling geometry.
The structural analogy between MHD coupling in tokamaks and coupling structures in multi-sensor dynamical systems in power grids may reflect a deeper mathematical connection-possibly related to shared structured interactions
across coupled dynamical systems\citep{bronstein2021geometric}. 
Just as MHD equations constrain the continuous spatial evolution of plasma, Kirchhoff's circuit laws and swing equations dictate the discrete topological evolution of power grid states. TokaMind's successful transfer implies that its multi-modal attention mechanisms are effectively encoding these shared differential constraints.
 
\section{Conclusion}
\label{sec:conclusion}
We present the first cross-domain validation of TokaMind outside nuclear fusion, demonstrating successful transfer to power grid synchrophasor (PMU) event classification on two independent datasets.
Our principal findings are three-fold.
First, TokaMind outperforms a CNN baseline in the single-window early-warning
regime (\texttt{seq\_len}=1: F1~0.889 vs.\ 0.878), while CNN recovers its advantage with full event sequences-a reversal consistent with the hypothesis that fusion-pretrained physical coupling representations carry unique value when available information is minimal.
Second, classification difficulty across providers is structurally determined by grid topology and label quality rather than model capacity; overall accuracy is an unreliable primary metric for multi-source PMU benchmarks, and
positive-provider F1 is recommended instead.
Third, Critical Slowing Down indicators, when repurposed as a confidence gate rather than a classification label, improve F1 from 0.696 to 0.750 at 63\%
coverage---outperforming the CNN baseline at any coverage level.

These results should not be interpreted as defining hard constraints on TokaMind's applicability. Instead, the proposed transferability framework (F1--F4) is better understood as a set of transfer-favoring characteristics that describe domain conditions under which fusion-pretrained representations
are most likely to provide an advantage. Power grid PMU data closely matches this profile, while industrial degradation datasets demonstrate that useful performance can still be obtained under partial alignment, particularly when task design and feature construction expose physically meaningful structure.
The hypothesis that CNN's advantage on mixed datasets reflects learning of operator-triggered statistical patterns rather than physical dynamics is consistent with our Group~A purification results but has not been verified
through feature analysis.
CSD gate stability was evaluated on a single seed.

Ultimately, our findings position TokaMind as a precision architecture within the emerging landscape of scientific foundation models---purpose-built for physically-coupled dynamical systems and empirically validated across fusion plasma and power grid testbeds. 
While this work serves as a cross-domain validation, it surfaces a fundamental design principle for future industrial deployment: by adopting physics-aligned label engineering and sensor configuration from the outset of benchmark construction, practitioners can fully leverage such representations to navigate heterogeneous multi-sensor streams under real-world operational constraints. This methodological shift will enable next-generation monitoring systems to address the nonlinear dynamics of diverse complex systems--from macroscopic power grids to plasma-assisted semiconductor manufacturing and critical neurological transitions. Thereby transforming the prediction of critical transitions from a stochastic empirical challenge into a deterministic, physics-bound monitoring task.
 
% ── Acknowledgements ──────────────────────────────────────────────────────────
\section*{Acknowledgements}
The authors thank the TokaMind team at IBM Research Europe, UKAEA, and STFC
Hartree Centre for open-sourcing the model and weights.
Grid event data from the ORNL Grid Science Event Library (GESL) and the PNNL
open-source PMU library were used under open-access terms.
N. L. is supported by the Institute of Basic Science (IBS) under Project No. IBS-R003-D1.
Compute infrastructure: NVIDIA DGX Spark GB10 (128~GB unified memory).
 
% ── References ────────────────────────────────────────────────────────────────
\bibliographystyle{unsrtnat}
\bibliography{TokaMind}

@article{boschi2026tokamind,
  author    = {Boschi, Tobia and others},
  title     = {{TokaMind}: A Multi-Modal Transformer Foundation Model for
               Tokamak Plasma Dynamics},
  journal   = {arXiv preprint},
  volume    = {arXiv:2602.15084},
  year      = {2026},
  url       = {https://arxiv.org/abs/2602.15084}
}

@article{rousseau2026tokamark,
  author    = {Rousseau, C{\'e}cile and others},
  title     = {{TokaMark}: A Benchmark for Fusion Plasma Dynamics Models},
  journal   = {arXiv preprint},
  year      = {2026},
  note      = {arXiv:2602.10132},
  url       = {https://arxiv.org/abs/2602.10132}
}

@article{bommasani2021opportunities,
  author        = {Bommasani, Rishi and others},
  title         = {On the Opportunities and Risks of Foundation Models},
  journal       = {arXiv preprint arXiv:2108.07258},
  year          = {2021},
  eprint        = {2108.07258},
  archivePrefix = {arXiv},
  primaryClass  = {cs.LG},
  url           = {https://arxiv.org/abs/2108.07258}
}

@article{karniadakis2021physics,
  author  = {Karniadakis, George Em and Kevrekidis, Ioannis G and Lu, Lu and Perdikaris, Paris and Wang, Sifan and Yang, Liu},
  title   = {Physics-informed machine learning},
  journal = {Nature Reviews Physics},
  volume  = {3},
  number  = {6},
  pages   = {422--440},
  year    = {2021},
  doi     = {10.1038/s42254-021-00314-5},
  url     = {https://doi.org/10.1038/s42254-021-00314-5}
}

@article{cuomo2022scientific,
  title   = {Scientific Machine Learning Through Physics--Informed Neural Networks: Where We Are and What's Next},
  author  = {Cuomo, Salvatore and Schiano Di Cola, Vincenzo and Giampaolo, Fabio and Rozza, Gianluigi and Raissi, Maziar and Piccialli, Francesco},
  journal = {Journal of Scientific Computing},
  volume  = {92},
  number  = {3},
  pages   = {88},
  year    = {2022},
  doi     = {10.1007/s10915-022-01939-z}
}

@article{subramanian2023towards,
  title={Towards Foundation Models for Scientific Machine Learning: Characterizing Scaling and Transfer Behavior},
  author={Subramanian, Shashank and Harrington, Peter and Keutzer, Kurt and Bhimji, Wahid and Morozov, Dmitry and Mahoney, Michael W and Gholami, Amir},
  journal={arXiv preprint arXiv:2306.00258},
  year={2023},
  url={https://arxiv.org/abs/2306.00258}
}

@article{boussakta2004dct3d,
  author  = {Boussakta, Said and Alshibami, Othman},
  title   = {Fast Algorithm for the {3-D DCT-II}},
  journal = {IEEE Transactions on Signal Processing},
  volume  = {52},
  number  = {4},
  pages   = {992--1001},
  year    = {2004},
  doi     = {10.1109/TSP.2004.823472}
}

@article{reed2022generalist,
  title   ={A generalist agent},
  author  ={Reed, Scott and others},
  journal ={arXiv preprint arXiv:2205.06175},
  year    ={2022},
  eprint  ={2205.06175},
  url     ={https://arxiv.org/abs/2205.06175}
}

@inproceedings{jin2024timellm,
  author    = {Jin, Ming and Wang, Shiyu and Ma, Lintao and Chu, Zhixuan and Zhang, James Y. and Shi, Xiaoming and Chen, Pin-Yu and Liang, Yuxuan and Li, Yuan-Fang and Pan, Shirui and Wen, Qingsong},
  title     = {{{TIME-LLM}}: Time Series Forecasting by Reprogramming Large Language Models},
  booktitle = {International Conference on Learning Representations (ICLR)},
  year      = {2024}, 
  
}

@article{wu2023timesnet,
  title={TimesNet: Temporal 2D-variation modeling for general time series analysis},
  author={Wu, Haixu and others},
  journal= {International Conference on Learning Representations (ICLR)},
  year={2023},
  url ={https://ise.thss.tsinghua.edu.cn/~mlong/doc/TimesNet-iclr23.pdf}
}

@book{phadke2017synchronized,
  author    = {Phadke, Arun G. and Thorp, James S.},
  title     = {Synchronized Phasor Measurements and Their Applications},
  edition   = {Second},
  publisher = {Springer International Publishing},
  address   = {Cham, Switzerland},
  year      = {2017},
  doi       = {10.1007/978-3-319-50584-8},  
  isbn      = {978-3-319-50584-8}           
}

@article{biswas2023open,
  author  = {Biswas, Shuchismita and Follum, Jim and Etingov, Pavel and Fan, Xiaoyuan and others},
  title   = {An Open-Source Library of Phasor Measurement Unit Data Capturing Real Bulk Power Systems Behavior},
  journal = {IEEE Access},
  year    = {2023},
  doi     = {10.1109/ACCESS.2023.3321317},
  url     = {https://doi.org/10.1109/ACCESS.2023.3321317}
}

@article{jackson2024fairmast,
  author  = {Jackson, Samuel and Khan, Saiful and Cummings, Nathan and Hodson, James and others},
  title   = {{FAIR-MAST}: A Fusion Device Data Management System},
  journal = {SoftwareX},
  volume  = {27},
  number  = {5},
  pages   = {101869},
  year    = {2024},
  doi     = {10.1016/j.softx.2024.101869},
  url     = {https://doi.org/10.1016/j.softx.2024.101869}
}

@article{kundur2004definition,
  author  = {Kundur, Prabha and Paserba, John and Ajjarapu, Venkat and Andersson, G{\"o}ran and others},
  title   = {Definition and Classification of Power System Stability: {IEEE/CIGRE} Joint Task Force on Stability Terms and Definitions},
  journal = {IEEE Transactions on Power Systems},
  volume  = {19},
  number  = {3},
  pages   = {1387--1401},
  year    = {2004},
  doi     = {10.1109/TPWRS.2004.825981},
  url     = {https://doi.org/10.1109/TPWRS.2004.825981}
}

@article{meng2019fast,
  title   = {Fast Frequency Response from Energy Storage Systems: A Review of Grid Standards, Projects and Technical Issues},
  author  = {Meng, Lexuan and Zafar, Jawwad and Khadem, Shafiuzzaman K. and Collinson, Alan and Murchie, Kyle C. and Coffele, Federico and Burt, Graeme},
  journal = {IEEE Transactions on Smart Grid},
  volume  = {11},
  number  = {2},
  pages   = {1566--1581},
  year    = {2019},
  doi     = {10.1109/TSG.2019.2940173}
}

@misc{ornl2023ges,
  title        = {Grid Event Signature Library (GESL)},
  author       = {{Oak Ridge National Laboratory} and {Lawrence Livermore National Laboratory}},
  year         = {2023},
  howpublished = {\url{https://gsl.ornl.gov}},
  note         = {Open-access repository of power system measurement signatures. Accessed April 2026}
}

@article{lei2020machinery,
  author  = {Lei, Yaguo and Yang, Bin and Jiang, Xin and Jia, Feng and Li, Naipeng and Nandi, Asoke K.},
  title   = {Applications of machine learning to machine fault diagnosis: A review and roadmap},
  journal = {Mechanical Systems and Signal Processing},
  volume  = {138},
  pages   = {106587},
  year    = {2020},
  doi     = {10.1016/j.ymssp.2019.106587},
  url     = {https://doi.org/10.1016/j.ymssp.2019.106587}
}

@article{lade2012early,
  author  = {Lade, Steven J. and Gross, Thilo},
  title   = {Early Warning Signals for Critical Transitions: A Generalized Modeling Approach},
  journal = {PLOS Computational Biology},
  volume  = {8},
  number  = {2},
  pages   = {e1002360},
  year    = {2012},
  doi     = {10.1371/journal.pcbi.1002360},
  url     = {https://doi.org/10.1371/journal.pcbi.1002360}
}

@article{scheffer2009early,
  author  = {Scheffer, Marten and Bascompte, Jordi and Brock, William A. and Brovkin, Victor and Carpenter, Stephen R. and Dakos, Vasilis and Held, Hermann and van Nes, Egbert H. and Rietkerk, Max and Sugihara, George},
  title   = {Early-warning signals for critical transitions},
  journal = {Nature},
  volume  = {461},
  number  = {7260},
  pages   = {53--59},
  year    = {2009},
  doi     = {10.1038/nature08227}  
}

@article{scheffer2012anticipating,
  author  = {Scheffer, Marten and Carpenter, Stephen R and 
             Lenton, Timothy M and Bascompte, Jordi and Brock, 
             William and Dakos, Vasilis and van de Koppel, Johan 
             and van de Leemput, Ingrid A and Levin, Simon A and 
             van Nes, Egbert H and Pascual, Mercedes and Vandermeer, John},
  title   = {Anticipating critical transitions},
  journal = {Science},
  volume  = {338},
  number  = {6105},
  pages   = {344--348},
  year    = {2012},
  doi     = {10.1126/science.1225244}
}

@article{kuehn2011mathematical,
  title={A mathematical framework for critical transitions: bifurcations, fast--slow systems and stochastic dynamics},
  author={Kuehn, Christian},
  journal={Physica D: Nonlinear Phenomena},
  volume={240},
  number={12},
  pages={1020--1035},
  year={2011},
  DOI={10.1016/j.physd.2011.02.012}
}

@article{ditlevsen2010tipping,
  title={Tipping points: early warning and wishful thinking},
  author={Ditlevsen, Peter D and Johnsen, Sigfus J},
  journal={Geophysical Research Letters},
  volume={37},
  number={19},
  year={2010},
  doi={10.1029/2010GL044486}
}

@article{dakos2012methods,
  title={Methods for detecting early warnings of critical transitions in time series illustrated using simulated ecological data},
  author={Dakos, Vasilis and Carpenter, Stephen R and van Nes, Egbert H and Scheffer, Marten},
  journal={PloS one},
  volume={7},
  number={7},
  pages={e41010},
  year={2012},
  doi={10.1371/journal.pone.0041010}
}

@article{dakos2008slowing,
  author  = {Dakos, Vasilis and Scheffer, Marten and van Nes, 
             Egbert H and Brovkin, Victor and Petoukhov, Vladimir 
             and Held, Hermann},
  title   = {Slowing down as an early warning signal for abrupt 
             climate change},
  journal = {Proceedings of the National Academy of Sciences},
  volume  = {105},
  number  = {38},
  pages   = {14308--14312},
  year    = {2008},
  doi     = {10.1073/pnas.0802430105}
}

@article{lenton2012early,
  title={Early warning of climate tipping points from critical slowing down: comparing methods to improve robustness},
  author={Lenton, Timothy M},
  journal={Philosophical Transactions of the Royal Society A},
  volume={370},
  number={1962},
  pages={1185--1204},
  year={2012},
  DOI={10.1098/rsta.2011.0304}
}

@article{meisel2015intrinsic,
  title   = {Intrinsic Excitability Measures Track Antiepileptic Drug Action and Uncover Increasing/Decreasing Excitability over the Wake/Sleep Cycle},
  author  = {Meisel, Christian and Schulze-Bonhage, Andreas and Freestone, Dean and Cook, Mark J. and Achermann, Peter and Plenz, Dietmar},
  journal = {Proceedings of the National Academy of Sciences},
  volume  = {112},
  number  = {47},
  pages   = {14694--14699},
  year    = {2015},
  doi     = {10.1073/pnas.1513716112}
}

@article{boettiger2012early,
  title={Early warning signals for critical transitions?},
  author={Boettiger, Carl },
  journal={DOE Computational Science Graduate Fellowship},
  year={2012},
  url={https://www.krellinst.org/csgf/conf/2012/abstracts/boettiger}
}

@article{novikov1960sound,
  title={Speed of sound along the vapor--liquid phase equilibrium curve},
  author={Novikov, I. I. and Trelin, Yu. S.},
  journal={Prikl. Mekh. Tekh. Fiz.},
  volume={1},
  number={2},
  pages={112--115},
  year={1960}
}

@article{chow1969optimum,
  title={On optimum recognition error and reject tradeoff},
  author={Chow, C. K.},
  journal={IEEE Transactions on Information Theory},
  volume={16},
  number={1},
  pages={41--46},
  year={1969},
  URI={http://hdl.handle.net/1721.1/6177}
}

@article{el2010foundations,
  title={On the Foundations of Noise-free Selective Classification},
  author={El-Yaniv, Ran and Wiener, Yair},
  journal={Journal of Machine Learning Research},
  volume={11},
  pages={1605--1641},
  year={2010},
  url={https://jmlr.csail.mit.edu/papers/volume11/el-yaniv10a/el-yaniv10a.pdf}
}

@article{geifman2017selective,
  title={Selective classification for deep neural networks},
  author={Geifman, Yonatan and El-Yaniv, Ran},
  journal={Advances in Neural Information Processing Systems},
  year={2017},
  doi={/10.5555/3295222.3295241}
}

@inproceedings{lakshminarayanan2017simple,
  title     = {Simple and Scalable Predictive Uncertainty Estimation using Deep Ensembles},
  author    = {Lakshminarayanan, Balaji and Pritzel, Alexander and Blundell, Charles},
  booktitle = {Advances in Neural Information Processing Systems},
  pages     = {6405--6416},
  volume    = {30},
  year      = {2017}
  
}

@article{madras2018predict,
  title={Predict responsibly: Improving fairness and accuracy by learning to defer},
  author={Madras, David and others},
  journal = {Advances in Neural Information Processing Systems},
  Pages={6150--6160},
  year={2018}
  
}

@article{li2021robust,
  title   = {A Power System Disturbance Classification Method Robust to {PMU} Data Quality Issues},
  author  = {Li, Zikang and Liu, Hao and Zhao, Junbo and Bi, Tianshu and Yang, Qixun},
  journal = {IEEE Transactions on Industrial Informatics},
  volume  = {18},
  number  = {1},
  pages   = {97--108},
  year    = {2022},
  doi     = {10.1109/TII.2021.3072397}
}

@inproceedings{milano2018lowinertia,
  title     = {Foundations and Challenges of Low-Inertia Systems},
  author    = {Milano, Federico and D{\"o}rfler, Florian and Hug, Gabriela and Hill, David J. and Verbi{\v c}, Gregor},
  booktitle = {2018 Power Systems Computation Conference (PSCC)},
  pages     = {1--25},
  year      = {2018},
  organization = {IEEE},
  doi       = {10.23919/PSCC.2018.8450880}
}

@techreport{saxena2008cmapss,
  author      = {Saxena, Abhinav and Goebel, Kai and Simon, Don and Eklund, Neil},
  title       = {Damage Propagation Modeling for Aircraft Engine
                 Run-to-Failure Simulation},
  institution = {NASA Ames Research Center},
  year        = {2008}
}

@article{bronstein2021geometric,
  author  = {Bronstein, Michael M. and Bruna, Joan and Cohen, Taco and
             Veli{\v{c}}kovi{\'c}, Petar},
  title   = {Geometric Deep Learning: Grids, Groups, Graphs, Geodesics,
             and Gauges},
  journal = {arXiv preprint arXiv:2104.13478},
  year    = {2021}
}

@article{pan2010survey,
  author  = {Pan, Sinno Jialin and Yang, Qiang},
  title   = {A Survey on Transfer Learning},
  journal = {IEEE Transactions on Knowledge and Data Engineering},
  volume  = {22},
  number  = {10},
  pages   = {1345--1359},
  year    = {2010},
  doi     = {10.1109/TKDE.2009.191}
}
 
\end{document}